\documentclass[natbib=false,manuscript]{acmart}
\setcopyright{acmlicensed}
\copyrightyear{2026}
\acmYear{2026}
\acmDOI{XXXXXXX.XXXXXXX}
\acmISBN{978-1-4503-XXXX-X/2018/06}

\usepackage{todonotes}
\usepackage{booktabs}
\usepackage{xcolor}
\usepackage{hyperref}
\usepackage[
backend=biber,
style=acmnumeric,
sorting=ynt
]{biblatex}
\begin{document}

\author{Neil A. Ernst}
\orcid{0000-0001-5992-2366}
\affiliation{
  \institution{University of Victoria}
  \country{Canada}
}
\email{nernst@uvic.ca}

\author{Christoph Treude}
\orcid{0000-0002-6919-2149}
\affiliation{
  \institution{Singapore Management University}
  \country{Singapore}
}
\email{ctreude@smu.edu.sg}

\title{GenAI Is No Silver Bullet for Qualitative Research in Software Engineering}

\begin{abstract}
Qualitative research gives rich insights into the quintessentially human aspects of software engineering as a socio-technical system.
Qualitative research spans diverse strategies and methods, from interpretivist, in situ observational field studies, to deductive coding of data from mining studies. 
Advances in large language models and generative AI (GenAI) have prompted claims that artificial intelligence could automate qualitative analysis. Such claims are overgeneralizing from narrow successes. 
GenAI support must be carefully adapted to the data of interest, but also to the characteristics of a particular research strategy. In this \textbf{Frontiers of SE paper}, we discuss the emerging use of GenAI in relation to the broad spectrum of qualitative research in software engineering. 
We 
outline the dimensions of qualitative work in software engineering, %
scan limited emerging empirical evidence for GenAI assistance, %
examine the promises and pitfalls of GenAI-assisted qualitative research, %
and revisit qualitative research quality factors, in light of GenAI. %
Our goal is to inform researchers about the promises and pitfalls of GenAI-assisted qualitative research. We conclude with a research agenda to advance understanding of its use in software engineering.
\end{abstract}



\maketitle

\section{Introduction}

Research in software engineering (SE) increasingly recognizes that people are the main factor~\cite{lenberg2024qualitative,seaman2025qualitative}. Qualitative research in SE employs a wide range of research strategies~\cite{storeyWhoWhatHow2020}, such as field studies and sample surveys, each of which may apply different research methods, including semi-structured interviews, observations, grounded theory, or phenomenological and narrative analyses~\cite{lenberg2024qualitative}. These strategies allow researchers to study phenomena in their natural context~\cite{runeson2009guidelines}, uncover emerging meanings, and build theory. 

A distinctive feature of qualitative research in SE is the complexity of its data sources. Researchers analyze not only heterogeneous artifacts, such as source code, version control logs, issue tracker comments, design documents, and chat logs, but also how these artifacts are interconnected in socio-technical workflows. In addition, such artifacts must usually be interpreted alongside human-centered data such as interviews, surveys, and field observations. This combination of technical, social, and organizational material complicates qualitative analysis, and demands that methods be carefully adapted to the realities of SE research. 

Qualitative analysis often includes coding, i.e., assigning researcher-derived labels to segments of data, but it is not exhausted by label assignment. Researchers must also prepare and correct transcripts, become familiar with the material, compare cases, refine or extend codebooks, notice negative cases, and develop themes or theory. For example, Hoda conducted 58 interviews for her work on Agile teams, each one hour in length~\cite{hoda2011self}. A conservative estimate is that one hour of interview will produce 20 pages of text\footnote{\cite{montes2025large} reports 177 single spaced 12 pt pages for \~900 minutes of interviews}. Thus this work required the researcher (after doing the interviews) to review, correct, and iteratively analyze over one thousand pages of transcripts. The outcome is not merely a labeled dataset, but a series of themes and, in some cases, a theory around the phenomena of interest. 

The advent of large language models and related GenAI technologies raises important questions for traditional approaches to qualitative research. GenAI models can summarize, translate, and classify text, and they have already been tested in qualitative coding tasks. For example, a recent requirements engineering study found that a state-of-the-art model achieved substantial agreement with human analysts for deductive annotation (Cohen's $\kappa$ > 0.7), but performed poorly in zero-shot settings (zero-shot describes a prompt that does not give an example of how the model should label the data). Carefully designed prompts improved both precision and reliability~\cite{shah2025inductive}. 
A mapping study of large language models in qualitative research similarly concluded that the current evidence supports only narrow annotation tasks, with prompt engineering and human supervision remaining essential, and with inaccuracies or hallucinations still possible~\cite{barros2025large}. 

Other work has shown that in some annotation tasks, large language models reach levels of agreement with human annotators comparable to inter-human agreement, suggesting their potential role as partial substitutes in evaluation studies~\cite{ahmed2025can}. In related fields, hybrid workflows often assign humans the task of developing the codebook and adapting it for the use of the model, while the model output is validated against a human gold standard~\cite{Blondeel2025,dunivin2025scaling}.

Since LLMs are a relatively recent technology, few primary studies have used them extensively (as we show in \S \ref{sec:slr}). In other fields, researchers have proposed frameworks~\cite{NguyenTrung2025}. Finally, major Qualitative Data Analysis (QDA) vendors, such as MaxQDA\footnote{https://www.maxqda.com/products/ai-assist} or Atlas.ti (tagline: ``Master Your Research Projects with the Power of AI''), have moved quickly to add AI capabilities to these tools. In environments that auto-accept AI suggestions, it may become impossible to disentangle AI from human labels. %

These trends highlight both the promise of GenAI and the risks of overstating its capabilities. Against this backdrop, this paper asks: Where can GenAI add value to qualitative SE research, and where does it fall short? We first characterize the diversity of qualitative SE research and the epistemological distinctions that shape acceptable forms of support. We then present a limited scan of some key publication venues. The scan is intended to motivate the research agenda, but cannot estimate prevalence. On that basis, we discuss promises and pitfalls, revisit qualitative research quality, and conclude with a research agenda for the SE community.

\section{The Spectrum of Qualitative Research in SE}

To understand where GenAI may or may not add value and what is currently supported by evidence, it is necessary first to consider the spectrum of research strategies and the underlying methods used in qualitative SE. These methods differ in their objectives, data requirements, and epistemological assumptions, which, in turn, shape the opportunities and limitations for GenAI support. We use the distinctions elaborated in Storey et al.~\cite{storeyWhoWhatHow2020} to guide the discussion. %

\textbf{Respondent Strategies} ask a sample of participants for feedback on tools, processes, or challenges. Responses can be elicited through methods such as interviews, surveys, or focus groups. Good practice includes careful preparation and transparent reporting~\cite{hove2005experiences}. For example, a study of secure SE interviewed practitioners in 11 companies to identify practices, mechanisms of knowledge sharing, and challenges~\cite{arora2021secure}. Surveys and questionnaires often complement interviews; open-ended questions analyzed thematically can also yield qualitative insights. These strategies maximize generalizability but may not be conducted in realistic or controlled situations. 

\textbf{Field Strategies} embed researchers into realistic settings, in order to observe practices and interactions directly. Although resource intensive, a field study provides unique insight into how software teams operate. A multi-company ethnography of DevOps and microservices adoption, for example, combined months of participation with follow-up interviews, revealing benefits such as rapid delivery, and challenges such as coordination overhead~\cite{dittrich2024teaching}.

\textbf{Lab Strategies} and \textbf{Data Strategies}, while most common in SE research, less commonly use qualitative research methods. Lab approaches emphasize control using experiment and simulation, and are predominantly focused on quantitative, post-positivist data analysis. Data strategies such as mining studies and experiments likewise analyze artifacts using quantitative approaches. Occasionally such strategies triangulate the experimental results by collecting free-form text responses in a survey or small-scale user study. These responses are then analyzed using some form of text analysis. 

To make sense of the data collected, common interpretive approaches include \textbf{grounded theory} and \textbf{thematic analysis}. Grounded theory builds theory through iterative coding and sampling, and has been widely applied in contexts with few prior hypotheses~\cite{seaman2025qualitative}. For example, earlier we mentioned Hoda's grounded theory study on agile self-organization. The study involved 58 professionals in 23 organizations, identifying roles, balancing practices, and contextual factors that support self-organization~\cite{hoda2011self}. Thematic analysis, in contrast, focuses on identifying and reporting patterns in data and can be carried out inductively or deductively~\cite{braunUsingThematicAnalysis2006}.
Recent commentaries encourage the greater use of interpretive methods, such as phenomenological, narrative, and discourse analysis, and stress the importance of reflecting on epistemological assumptions before choosing a method~\cite{lenberg2024qualitative}.

\subsection{Dimensions of Qualitative Research}
\noindent\textbf{Epistemological and ontological orientation.} 
A researcher's epistemological and ontological stances, \emph{often unconscious,} shape subsequent research strategy and method choices~\cite{Easterbrook2008}. Epistemology concerns what counts as knowledge and how that knowledge is produced. Ontology concerns what exists in the world and how one thinks about reality. Following Easterbrook et al.~\cite{Easterbrook2008} (itself derived from Creswell~\cite{creswell2002qualitative}), we use terms such as \emph{positivist}, \emph{post-positivist}, \emph{constructivist}, and \emph{interpretivist} as practical labels rather than fixed boxes. 

In this paper, \emph{positivist} and \emph{post-positivist} approaches treat coding outcomes as observations to be stabilized, compared, and evaluated, while \emph{constructivist} and \emph{interpretivist} approaches\footnote{These terms are used interchangeably in the literature and are complementary, but constructivism is an ontological orientation while interpretivism is an epistemological one.} emphasize findings as situated and co-produced through engagement among researchers, participants, artifacts, and theory~\cite{lenberg2024qualitative}. These labels are contested, often entangled with ontological commitments, and do not map neatly one-to-one onto methods~\cite{sharp2016ethnographic}. 

For that reason, the relevant question is not whether a method is ``an AI method,'' but which forms of GenAI support remain compatible with a study's epistemological commitments. Narrow autonomous support is easiest to defend in highly deductive, codebook-driven tasks; it is much harder to defend in reflexive thematic analysis, grounded theory, or ethnography, where meaning is developed through situated interpretation and reflexivity~\cite{Jowsey2025,braunUsingThematicAnalysis2006}. A related distinction in social sciences is called Big-Q (reflexive/constructivist) vs little-q (post-positivist)~\cite{Braun2024}.

\noindent\textbf{Coding strategy.} 
Coding assigns labels to segments of data, but qualitative analysis also involves memoing, comparing cases, refining concepts, and deciding when apparent exceptions should force reinterpretation. Inductive approaches, such as grounded theory and some forms of thematic analysis, develop categories through engagement with the data. Deductive approaches apply predefined codebooks, for example from prior theory or evaluation criteria. Hybrid strategies that move between inductive discovery and deductive comparison are also common. A single negative case can matter disproportionately because it may expose a mistaken explanation rather than simply add another coded excerpt~\cite{unphon2010architecture}.

\noindent\textbf{Data granularity and type.} 
Units of analysis can range from fine-grained (e.g., message-level coding of issue tracker comments) to coarse-grained (e.g., entire interview narratives). Sources include textual artifacts, socio-technical traces, and multimodal materials.

\noindent\textbf{Iteration and researcher roles.} 
Iteration, i.e., the repeated revisiting of data and analysis, is central for qualitative research but varies across methods. Grounded theory cycles between data collection and analysis, while thematic analysis refines codes over multiple passes. Some studies involve multiple coders, not only to distribute labor or estimate agreement, but also to surface disagreement, extend the codebook, and identify negative cases. Ethnographies, by contrast, often rely on prolonged engagement by a single researcher who must connect interviews, observations, and artifacts over time~\cite{sharp2016ethnographic}. Reflexivity, i.e., researchers' critical reflection on how their own background and assumptions shape the study, is essential for constructivist methods~\cite{lenberg2024qualitative,braunUsingThematicAnalysis2006}. Some of this sensemaking depends on what Eriks{\'e}n called \emph{inverted indexicality}: traces of artifacts, actors, or routines outside the immediate excerpt that are not present in the data fragment itself, but are necessary to interpret what is going on~\cite{eriksen1998knowing}.

Qualitative research in SE is diverse and multi-hued: there is variety in research strategies and underlying research methods and data; variety in the analytic techniques used; variety in epistemological and researcher roles; and variety in the sensemaking and coding practices. Any GenAI support must therefore be tailored to specific tasks and contexts rather than assumed to generalize across all of them.

\section{GenAI for Qualitative SE Research}

Empirical evidence of the impact of GenAI assistance in qualitative SE research is currently limited and highly context-dependent. By far the biggest impact may be in automating transcription of audio files, using tools such as Whisper~\cite{pmlr-v202-radford23a}. However, while incredibly useful, transcription is a limited and relatively uncontroversial use.\footnote{``relatively"; transcription inherently involves value judgments in deciding what is being said and what to leave out~\cite{Poland1995}, privileges speakers with similar speech patterns to training data, etc. } 

\subsection{Current Usage: An Illustrative Venue Scan}
\label{sec:slr}
To ground the discussion in visible reporting practice, we conducted a limited venue scan of recent publications. The purpose of this scan is to motivate the research agenda with concrete examples, not to estimate prevalence or claim systematic coverage. We therefore treat it as illustrative context rather than review evidence.

We selected three 2025 conference proceedings in which qualitative work is common in SE or adjacent communities: ICSE 2025 (n=258), CHASE 2025 (n=37), and CSCW 2025 (strictly speaking, PACMHCI vol. 9, CSCW2, n=312). We focused on conferences because journals typically lag conference publication cycles, and we excluded arXiv because it is not peer reviewed. Table \ref{tab:paper_review} reports the filtering steps and counts. Our replication package is located at \url{https://doi.org/10.5281/zenodo.21631436}, and contains tables listing all papers, filtered papers, and scripts for replication. The relatively small amount of qualitative papers at ICSE compared with the other venues likely reflects a broader focus on strictly quantitative/tool development papers at ICSE.

Our screening proceeded in three steps. First, we screened in studies that used qualitative coding by searching the PDFs for the keywords `cod[ed|ing]' and `qualitative'.
Second, we prompted ChatGPT (December 2025) to read each screened-in paper and label it \emph{yes}, \emph{no}, or \emph{maybe} with respect to using an LLM as an annotator, requiring a supporting quotation and location for every \emph{yes} or \emph{maybe} label; one author then read each flagged passage and resolved the \emph{maybe} labels manually. The full prompt is in the replication package.

Third, for papers that matched the screening criteria but did not use GenAI, we investigated their use of qualitative data analysis (QDA) software in order to understand what was occurring: one author read the Method section of each paper to see whether the authors reported tool support for the coding process, using the list of QDA software from Wikipedia plus Google Sheets~\footnote{Google, Aquad, Atlas.ti, Cassandre, CLAN, Coding Analysis Toolkit (CAT), Compendium, Dovetail, Dedoose, ELAN, KH Coder, MAXQDA, NVivo, QDAcity, QDA Miner, Qiqqa, Quantitative Discourse Analysis Package (qdap) (R package), Quirkos, RQDA (R package), Taguette, Transana, XSight}.

\subsubsection{Results}
Within this limited scan, 20 of the 209 screened-in CSCW papers (10\%), five of 25 at ICSE (20\%), and two of 16 at CHASE (13\%) reported using any tool to support coding. Seven CSCW papers (3\% of its screened-in papers) substantively reported GenAI use for qualitative analysis; no ICSE or CHASE paper did. The remaining papers reported human coding and labeling approaches. Table~\ref{tab:paper_review} reports the complete counts.

Of the seven CSCW papers, three are meta-studies looking at how LLMs would be used by researchers, for example, in building a visualization tool to help theme development with GenAI~\cite{Kang2025}. Three applied GenAI in positivist/deductive annotation tasks, for example by applying an existing labeling framework to datasets~\cite{10.1145/3757592}. The final paper used LLMs to simplify and scale data extraction by selecting a subset of videos~\cite{10.1145/3757697}. In all CSCW cases, it is striking that LLMs are not used in an \textbf{inductive} thematic-analysis sense, although the three meta-LLM papers all suggest this development is expected.

Absent phrasing such as ``we \emph{manually} coded the data,'' however, it is not possible to definitively determine GenAI use. For example, \cite{you_atlas} reported ``we utilized the computer-assisted qualitative data analysis software Atlas.ti to facilitate our coding process.'' The latest qualitative data analysis tools like MaxQDA and Atlas.ti now have GenAI support; even Google Sheets offers AI-assisted autocomplete. Thus, reported tool use does not cleanly separate human-only and AI-assisted coding, and may hide low-visibility forms of assistance. As we recommend below, future papers should disclose GenAI use, even when it is limited to autocomplete in a spreadsheet.

These counts should be interpreted cautiously. They indicate visible reporting in this sample, not broader prevalence in SE research. CSCW proceedings cover a more recent publication date than ICSE or CHASE: the notification deadline for CSCW 2025 was in multiple stages, with the latest being August 6, 2025, whereas for ICSE/CHASE 2025 it was January 22, 2025, nearly six months earlier. This, together with the far larger number of qualitative papers published at CSCW, is a plausible explanation for the observed variance in reported use.

\begin{table}[h]
    \centering
    \caption{Summary of Paper Review and Filtering. Percentages for qualitative coding papers are of all papers in the proceedings; percentages for tool and GenAI use are of the qualitative coding papers at that venue.}
    \label{tab:paper_review}
    \begin{tabular}{lccc}
        \toprule
        Stage & ICSE 2025 & CHASE 2025 & CSCW 2025 \\
        \midrule
        Overall Papers & 258 & 37 & 312 \\
        Qualitative Coding Papers & 25 (10\%) & 16 (43\%) & 209 (67\%) \\
        Reported Tool Use for Coding & 5 (20\%) & 2 (13\%) & 20 (10\%) \\
        Reported GenAI Use for Coding & 0 (0\%) & 0 (0\%) & 7 (3\%) \\
        \bottomrule
    \end{tabular}
\end{table}

\subsection{Reported Uses Beyond the Venue Scan}
Studies of GenAI for qualitative SE research have so far concentrated on a small set of assistive uses: (i) \textit{deductive coding and annotation}, where large language models act as additional coders against an existing codebook; (ii) \textit{summarization and translation}, where models condense or translate textual data; and (iii) \textit{conceptual support}, where models surface candidate codes, clusters, or patterns for researcher review. These assistive uses are useful but narrow. 

Most of the supporting evidence concerns annotation tasks in deductive studies, short text snippets, or human-in-the-loop workflows where researchers retain responsibility for interpretation. That pattern is consistent with the venue scan above: the CSCW papers that visibly report GenAI use focus on annotation, data reduction, or tooling around analysis rather than autonomous inductive theme development. Evidence is thin or non-existent for interpretivist epistemologies, inductive theme development, grounded theory, ethnography, and other forms of qualitative work in which interpretation is itself the central analytic contribution.

Ahmed et al.~evaluated six state-of-the-art LLMs across ten annotation tasks and five datasets. For deductive coding tasks that required little contextual awareness, such as code summarization or checking name-value consistency, the models reached human-level inter-rater agreement. In contrast, for tasks that depended heavily on context, such as inferring causality or interpreting static analysis warnings, performance was substantially lower~\cite{ahmed2025can}. These results suggest that GenAI can complement deductive coding with well-defined labels but struggles when success depends on interpreting relationships, dependencies, or higher-level meaning.

In requirements engineering, Shah et al.~tested a large language model as a second annotator for the deductive coding of user stories. The model achieved substantial agreement with human analysts (Cohen's $\kappa$ > 0.7) and delivered consistent results in all runs~\cite{shah2025inductive}. By contrast, zero-shot prompting performed poorly, underscoring the need for human-prepared codebooks. This is an example of a positive human-AI collaboration case: people define the construct, prepare examples, and review outputs, while the model contributes speed and consistency on a constrained labeling task. A mapping study confirmed these findings, concluding that the evidence so far covers only narrow tasks such as sentiment analysis, translation, and summarization, while interpretive tasks such as theme development remain outside the scope of current GenAI capabilities~\cite{barros2025large}.

A thematic analysis of expert essays on large language models in software development explored perceptions of quality, collaboration, and security~\cite{yadav2025thematic}. While sentiment analysis suggested general optimism, manual thematic coding revealed more nuanced trade-offs that automated keyword extraction tools failed to capture. This reinforces that unsupervised mappings are insufficient for thematic analysis, where context and nuance matter. It also illustrates a narrower assistive role for automation: summarization and sentiment-style screening can help researchers triage large corpora before interpretive analysis begins.

Closely related to this reflection paper is the recent work of Montes et al.~\cite{montes2025large}. They take existing interview transcripts coded with reflexive thematic analysis~\cite{braunUsingThematicAnalysis2006}, engineer prompts for GenAI, and compare the codes of the humans and GenAI in a blinded trial. There was a slight preference for the GenAI codes (61\%), but themes from GenAI were less aware of latent interpretations. The paper takes a post-positivist slant on the topic and does not interrogate the use of GenAI with respect to constructivist epistemologies.

Outside of SE, computational grounded theory illustrates how GenAI can scale qualitative work while keeping humans in control. Algorithms surface candidate topics and patterns, but interpretation and theory building remain human-led~\cite{alqazlan2025novel}. Similarly, Chen et al.~introduced computational metrics (coverage, density, novelty, divergence) to evaluate coding outputs, treating GenAI as a member of a coding team rather than a replacement~\cite{chen2024computational}. Hybrid workflows such as those proposed by Dunivin et al.~demonstrate how human-developed codebooks can be adapted for large language models, with outputs compared to gold standards to preserve interpretive depth~\cite{dunivin2025scaling}. 

The strongest collaboration pattern visible in the literature is one where researchers are responsible for theory, codebook design, disagreement resolution, and final interpretation, while GenAI helps with scale, candidate pattern surfacing, or secondary coding. Dai, Xiong, and Ku~\cite{dai-etal-2023-llm} advocate a human-in-the-loop model in applying LLMs to thematic analysis in a deductive/positivist approach (an approach fundamentally inconsistent with the underlying reflexive philosophy Braun and Clarke~\cite{braunUsingThematicAnalysis2006} espouse). GenAI improved efficiency but had significant disagreement with the human coder. Wen et al.~\cite{Wen2025} similarly report improved efficiency, but also significant concerns around contextual understanding and overly broad themes from the GenAI tool. 

Overall, the evidence base is narrow: successes mostly involve deductive coding and annotation of isolated statements, with limited and cautious experimentation in thematic and grounded-theory contexts. Ethnography and long-term case studies have remained unaffected by GenAI support to date. Importantly, most current findings apply to low-context tasks, whereas qualitative SE methods often require connecting multiple socio-technical artifacts and integrating them with human-centered data.

\section{Promises and Pitfalls of Using GenAI in Qualitative Research}
To highlight the implications, we now discuss the main promises and pitfalls of GenAI for qualitative SE research.

\subsection{Promises}
GenAI and large language models (LLMs) offer several promising avenues for qualitative research in software engineering, but the strongest evidence concerns assistive use in bounded tasks rather than open-ended interpretation.

First, LLMs can accelerate \textbf{deductive coding and annotation}. When a predefined codebook is available, LLMs can act as additional coders, quickly labeling large datasets and highlighting disagreements among coders~\cite{shah2025inductive}. Ahmed et al.~similarly found human-level agreement in several low-context annotation tasks, while performance degraded on tasks requiring stronger contextual inference~\cite{ahmed2025can}. Codebook-driven studies can use LLMs to scale screening and secondary annotation when labels are stable and the unit of analysis is narrow.

Second, LLMs enable \textbf{rapid summarization and translation}. The mapping study by Barros et al.~identifies these as among the most common supported uses~\cite{barros2025large}. In practice, this can help researchers triage long transcripts, multilingual corpora, or large sets of issue discussions before more interpretive analysis begins. The benefit is clearest when the researcher treats the model output as a compressed intermediate artifact to inspect and revise, not as a final interpretation.

Third, LLMs can provide \textbf{candidate codes, patterns, or comparisons for researcher review}. Computational grounded theory and related hybrid workflows use algorithms to surface candidate topics, clusters, or relationships while leaving theory building and interpretation to human analysts~\cite{alqazlan2025novel,dunivin2025scaling}. This kind of support is valuable when it helps researchers inspect a larger space of possibilities or compare alternative codings, especially in early-stage exploration.

Fourth, LLMs may support \textbf{larger and more heterogeneous datasets}, particularly artifact-centered SE studies. The current literature suggests value in screening issue discussions, developer chat, or other textual traces that would be difficult to review exhaustively by hand~\cite{alqazlan2025novel}. As the venue scan suggests, one realistic near-term benefit may be data reduction or prioritization rather than full analytic delegation.

Finally, GenAI can support \textbf{evaluation of coding outputs}. Metrics such as coverage, density, novelty, and divergence provide ways to compare human and GenAI coders in deductive or hybrid designs~\cite{chen2024computational}. This is useful not because agreement alone settles interpretive quality, but because it makes the limits of GenAI assistance more visible in the kinds of coder-like roles where the evidence is currently strongest.

\subsection{Pitfalls}
GenAI assistance in qualitative research comes with several potential pitfalls. One major concern is \textbf{overgeneralisation from narrow tasks and data}. The positive results discussed in Section~\ref{sec:slr} mostly concern annotation, summarization, or human-in-the-loop support on bounded inputs. There is limited support for GenAI in theme synthesis, theory building, or ethnography, and the existing findings do not readily generalize across the heterogeneous data types common in software engineering, such as source code, version control traces, issue discussions, design documents, and interview transcripts. Moreover, existing evidence largely examines whether GenAI can approximate the judgments of a single human coder in tightly scoped settings; it remains unclear whether GenAI can meaningfully substitute for multiple coders, or support consensus-building roles, which are central to many qualitative analysis practices. The current evidence base should therefore not be misinterpreted as GenAI \emph{solving} qualitative research~\cite{barros2025large} (nor that this is a desired goal). 

Another limitation is the \textbf{lack of context and interpretive depth}. LLMs excel at pattern recognition, but they lack socially embedded sensemaking~\cite{alqazlan2025novel}. This means they are poorly positioned to notice negative cases that should force analysts to revise an emerging explanation rather than merely add another label~\cite{unphon2010architecture}. They are also unlikely to detect what Eriks{\'e}n calls \emph{inverted indexicality}: traces of artifacts, actors, or routines outside the immediate excerpt that are not present in the data fragment itself, but are crucial for interpretation~\cite{eriksen1998knowing}. 

This limitation matters most for approaches such as grounded theory, ethnography, or reflexive thematic analysis, where researchers interpret interconnected socio-technical artifacts and lived experiences together, often in dialogue with prior theory, and where the researcher reflects on their positionality~\cite{sharp2016ethnographic,DeSousa2025}. In these settings overall efficiency (e.g., number of transcripts parsed, or number of codes generated) is not the limiting factor. 

A further risk involves \textbf{bias and hallucinations}. GenAI systems inherit biases from their training data and may hallucinate plausible but incorrect codes or summaries. Humans of course also exhibit bias, but mechanisms such as reflexivity and positionality statements at least make that possible bias visible. Hallucinations and hidden bias are especially problematic when researchers rely on model outputs without inspecting the underlying data, or when the study examines novel contexts that are not represented in training corpora~\cite{chen2024computational}. 

Additionally, there is the issue of \textbf{prompt sensitivity and reproducibility}. Model outputs depend heavily on the wording of prompts and random seeds, which can lead to inconsistent results across runs~\cite{shah2025inductive}. This undermines reliability, particularly in deductive studies where intercoder agreement is important.

Finally, researchers must consider the \textbf{epistemological mismatch} between GenAI and constructivist approaches. Constructivist methods emphasize the co-construction of meaning with participants, and treating GenAI as an ``additional coder'' can conflict with this orientation, especially when theory building and researcher interpretation are central rather than ancillary~\cite{lenberg2024qualitative,Jowsey2025,DeSousa2025}.

\section{Revisiting Qualitative Research Quality}

Qualitative research quality begins with the creation of high quality data~\cite{Malterud2016}. This means, \emph{inter alia}, well-motivated sampling or case study selection, carefully designed and evaluated instruments, such as interview questionnaires, and attentive data curation and cleaning.

Long-standing qualitative research criteria such as reliability, validity, reflexivity, and ethical responsibility need to be reconsidered as GenAI becomes part of the process.\footnote{It should be noted that some researchers question the idea that standards to ensure rigor are even relevant to qualitative research~\cite{sandelowski}.} 

\emph{Reliability} is often emphasized in SE because many studies use small and heterogeneous samples. 
Metrics such as Krippendorff's $\alpha$ 
apply only to annotation-style studies (e.g., surveys with deductive coding), not to interpretive methods such as grounded theory. When GenAI is introduced as a coder, researchers must still evaluate the agreement with humans and over repeated runs~\cite{shah2025inductive}. Complementary metrics such as coverage, density, and divergence~\cite{chen2024computational} can quantify differences.
These reliability statistics alone are insufficient without reflexive discussion and \emph{meaningful} member checking~\cite{Birt2016} with study participants such as developers. This caution follows directly from Section~\ref{sec:slr}: the strongest current evidence is for coder-like tasks with predefined labels, not for the broader interpretive practices that many qualitative SE studies rely on.

\emph{Validity} in qualitative SE research means ensuring that interpretations of developer artifacts, such as chat logs, design documents, or version control traces, credibly reflect practice. Section~\ref{sec:slr} suggested that summarization and secondary annotation are the most plausible current uses of GenAI. Even in those roles, the validity of GenAI annotations or codes can be threatened by hidden biases, misrepresented developer voices, or hallucinated results~\cite{chen2024computational}. Researchers must therefore remain reflexive about how GenAI shapes interpretation, and be transparent in documenting prompts, model versions, and parameter settings when analyzing SE artifacts~\cite{lenberg2024qualitative,baltes2026guidelinesempiricalstudiessoftware}. Positionality statements~\cite{DeSousa2025} can be an effective way to convey this to readers, especially in theory-oriented and interpretivist work where the researcher's stance is part of the analytic account.

\emph{Ethical and governance} issues are crucial in SE contexts because artifacts often contain sensitive or proprietary information. Using external GenAI services to process source code or internal communication logs raises confidentiality and intellectual property concerns~\cite{davison2024ethics}. Fairness is also critical: models trained mainly in open-source projects may underrepresent commercial or marginalized communities~\cite{treude2023she}. While this issue crosses research approaches, addressing these issues requires methodological safeguards and institutional guidance tailored to the mix of open and proprietary data from SE.

The role of GenAI must be kept in perspective. Across both the venue scan and the broader literature in Section~\ref{sec:slr}, current evidence shows that GenAI is useful mainly as a coding or summarization aid (e.g., labeling user stories or summarizing pull request discussions), not as an autonomous qualitative researcher. It lacks the epistemological grounding, reflexivity, and ethical responsibility expected of human researchers~\cite{lenberg2024qualitative,braunUsingThematicAnalysis2006}. Treating GenAI as a collaborator risks overstating its capabilities. For current practice, responsible use therefore means disclosing the exact tool and model used (cf. \cite{baltes2026guidelinesempiricalstudiessoftware}), distinguishing assistive roles from analytic roles, and justifying why that use fits the method's epistemology and theoretical commitments.
GenAI tools are better understood as tools that can accelerate some artifact-focused tasks, but still require careful human oversight~\cite{montes2025large}. 

\section{A Research Agenda}
Given the promises and perils, we highlight five directions for further work. Our aim is to leverage the capabilities GenAI may bring, while not losing the rich and subtle insights qualitative research approaches have brought to the community~\cite{seaman2025qualitative}.

\paragraph{Benchmarking GenAI in coder-like roles}
We must evaluate whether GenAI can substitute for additional coders in deductive coding or content analysis of socio-technical artifacts. Existing work has only examined whether LLMs can replace a single annotator~\cite{ahmed2025can} or in a single type of artifact~\cite{montes2025large}. Future studies should build on the work of Montes et al.~\cite{montes2025large} and compare human-human, human-AI, and AI-AI settings across a variety of artifacts such as source code, issue tracker comments, and commit messages~\cite{ahmed2025can,shah2025inductive}. This agenda should also ask which humans constitute the relevant benchmark, and whether agreement is always the right target, or whether critique, triangulation, or selective usefulness are more meaningful evaluation goals. The results so far suggest that LLMs perform best when tasks have well-established coding guides and require little contextual awareness, such as labeling isolated statements. In contrast, SE studies are deeply contextual.
Although narrow, this line of work provides a necessary baseline for assessing the reliability and contextual limits of GenAI in the simplest qualitative SE research tasks.

\paragraph{Extending GenAI evaluation to interpretive methods}
More ambitious research should examine how GenAI performs in interpretive methods such as grounded theory, field studies, and narrative analysis when applied to SE data. This is the main gap left by the current evidence base: Section~\ref{sec:slr} found some careful experimentation with thematic or grounded-theory-adjacent workflows, but little support for sustained interpretive sensemaking. For example, do model-proposed categories in grounded-theory studies of agile practices match human interpretations of that same data? Can GenAI summarize recurrent patterns in ethnographic field notes on DevOps teams? Nguyen and Welch~\cite{Nguyen2025} report, in the field of Organizational Behavior, that while AI was useful at finding specifics, it was poor at generating useful high-level summaries and required extensive manual oversight. Similarly, Montes et al.~\cite{montes2025large} report that GenAI was preferred in some situations, but in others lost context and nuance.

Addressing this gap requires not only mapping the role of GenAI in sensemaking, but also systematically comparing the processes and findings of human-only and GenAI-assisted research to understand the implications of involving GenAI in interpretive work.

\paragraph{Designing collaborative human-AI workflows}
Because GenAI-only analysis is generally insufficient, a challenge is to design workflows in which researchers and GenAI collaborate without losing interpretive depth. Section~\ref{sec:slr} suggests that the most promising pattern is not replacement, but staged collaboration: humans define the codebook or analytic frame, GenAI helps with screening or secondary coding, and humans then inspect disagreements, negative cases, and theory-relevant links. 

Future work should investigate what such collaborative workflows look like in practice, e.g., how researchers iterate between GenAI suggestions and their own coding, or how reflexive notes can be integrated into mixed human-AI analyses. 
Tool support will be essential: Interfaces should allow researchers to navigate interconnected artifacts, inspect model reasoning, support and record reflexive practices, and move seamlessly between levels of detail. 

\paragraph{Developing reporting and evaluation norms for SE research practice}
The SE research community must clarify how GenAI use should be documented and evaluated across study types without prematurely hardening those expectations into rigid standards. The venue scan suggests that current disclosure is often too coarse to distinguish human-only coding from AI-assisted tool use. Questions include: What level of disclosure is needed for qualitative studies using QDA tools with AI features? How should researchers distinguish assistive use from analytic delegation? How do we avoid documentation norms being used to exclude human-only interpretations?
Answers may build on emerging evaluation guidance for empirical SE studies with LLMs~\cite{baltes2026guidelinesempiricalstudiessoftware}, but will need to be adapted to the particular challenges of qualitative research with interconnected socio-technical artifacts.

\paragraph{Reconciling GenAI and constructivist and interpretivist research paradigms} Finally, it is worth asking under what conditions, if any, the use of GenAI can be reconciled with constructivist research paradigms. Notions of rigor in research are often derived from positivist philosophies, and ignore that some researchers view truth as a socially constructed phenomenon~\cite{sandelowski}. Results necessarily are filtered through a researcher and their biases, positionality, and theoretical lenses~\cite{DeSousa2025,sharp2016ethnographic}. What does that mean in the LLM era? As the seminal work by Kidder and Fine \cite{kidderQualitativeQuantitativeMethods1987} distinguishes, some research looks for answers, while other research looks for questions. GenAI would seem more suited to the former than the latter. The key question is therefore not whether GenAI can be inserted into any qualitative workflow, but which uses remain methodologically and epistemologically compatible without displacing the creativity, nuance, and subtlety that skilled qualitative researchers bring.

\begin{acks}
We are grateful to Martin Robillard, Umit Akirmak, and the anonymous reviewers for their insights and discussions around this topic.
\end{acks}

\printbibliography

\end{document}